\documentclass[final,5p,times,twocolumn]{elsarticle}

\usepackage{amssymb}
\usepackage{amsmath,bm}
\usepackage{caption}
\usepackage{graphicx}
\usepackage{subfigure}
\usepackage{hyperref}
\usepackage{indentfirst}
\usepackage{amsfonts}
\biboptions{numbers,sort&compress}
\usepackage{nomencl}
\makenomenclature
\usepackage{makeidx}
\usepackage{framed} 
\usepackage{multicol} 
\usepackage{nomencl} 
\usepackage{ifthen}
\makenomenclature
\renewcommand*\nompreamble{\begin{multicols}{2}}
\renewcommand*\nompostamble{\end{multicols}}

\RequirePackage{ifthen}
\renewcommand{\nomgroup}[1]{%
\ifthenelse{\equal{#1}{S}}{\item[\textbf{Subscript}]}{%
\ifthenelse{\equal{#1}{G}}{\item[\textbf{Greek}]}{}}}

\usepackage{lineno}

\begin{document}
\begin{frontmatter}


\title{Controlling and maximizing effective thermal properties by manipulating transient behaviors during energy-system cycles}


\author[mymainaddress]{Z. J. Gao}


\author[mymainaddress,mysecondaryaddress]{T. M. Shih\corref{mycorrespondingauthor}}
\ead{tmshih@xmu.edu.cn}

\author[mymainaddress,mythirdaryaddress]{H. Merlitz}

\author[myfourthlyaddress]{P. J. Pagni}

\author[myfifthaddress]{Z. Chen\corref{mycorrespondingauthor}}
\cortext[mycorrespondingauthor]{Corresponding author}
\ead{chenz@xmu.edu.cn}

\address[mymainaddress]{Department of Physics, Xiamen University, Xiamen, China 361005}
\address[mysecondaryaddress]{Institute for Complex Adaptive Matter, University of California, Davis, CA 95616, USA}
\address[mythirdaryaddress]{Leibniz Institute for Polymer Research, Dresden, Germany}
\address[myfourthlyaddress]{Department of Mechanical Engineering, University of California, Berkeley, CA 94720, USA}
\address[myfifthaddress]{Department of Electronic Science, Fujian Engineering Research Center for Solid-state Lighting,
State Key Laboratory for Physical Chemistry of Solid Surfaces, Xiamen University, Xiamen,
China 361005}

\begin{abstract}
Transient processes generally constitute part of energy-system cycles. If skillfully manipulated, they actually are capable of assisting systems to behave beneficially to suit designers' needs. In the present study, behaviors related to both thermal conductivities ($\kappa$) and heat capacities ($c_{v}$) are analyzed. Along with solutions of the temperature and the flow velocity obtained by means of theories and simulations, three findings are reported herein: $(1)$ effective $\kappa$ and effective $c_{v}$ can be controlled to vary from their intrinsic material-property values to a few orders of magnitude larger; $(2)$ a parameter, tentatively named as ``nonlinear thermal bias'', is identified and can be used as a criterion in estimating energies transferred into the system during heating processes and effective operating ranges of system temperatures; $(3)$ When a body of water, such as the immense ocean, is subject to the boundary condition of cold bottom and hot top, it may be feasible to manipulate transient behaviors of a solid propeller-like system such that the system can be turned by a weak buoyancy force, induced by the top-to-bottom heat conduction through the propeller, provided that the density of the propeller is selected to be close to that of the water. Such a turning motion serves both purposes of performing the hydraulic work and increasing the effective thermal conductivity of the system.
\end{abstract}


\end{frontmatter}


\section{Introduction}
\label{INTRODUCTION}
\begin{table*}[!t]
  \begin{framed}
    \printnomenclature
  \end{framed}
\end{table*}
Heat conduction and convection are mechanisms that govern thermal behaviors of various energy-related devices, including bi-segment thermal rectifiers with the thermal conductivity of the system depending on the temperature \cite{kobayashi2009oxide,chang2006solid}, a thermal diode model coupling two nonlinear $1D$ lattices for a wide range of system parameters \cite{li2004thermal}, thermoelectric modules with thermal energy being converted into electricity \cite{mahan2008thermoelectric,majumdar2004thermoelectricity}, low-temperature waste heat thermoelectric generator systems optimized and modified \cite{gou2010modeling}, photovoltaic films with solar energy being converted into electricity \cite{nelson2002organic,chow2010review,li2005high}, and light-emitting diodes with the electricity being converted into both thermal energy and the light \cite{schubert2005light,gessmann2004high}. On the basis of the first law of thermodynamics, the temperature of these solid energy systems is governed by (see Appendix A-$1$)
\begin{equation}\label{1}
\kappa\nabla^{2}T=\rho c_{v}\frac{\partial T}{\partial t}-\left(\frac{\partial\kappa}{\partial T}\right)\left[\left(\frac{\partial T}{\partial x}\right)^{2}+\left(\frac{\partial T}{\partial y}\right)^{2}+\left(\frac{\partial T}{\partial z}\right)^{2}\right]-Q_{g},
\end{equation}
\nomenclature[G]{$\kappa$}{thermal conductivity ($W\,m^{-1}\,K^{-1}$)}
\nomenclature[G]{$\rho$}{density ($Kg\,m^{-3}$)}%
\nomenclature{$c_{v}$}{heat capacity with volume kept constant ($Jkg^{-1}K^{-1}$)}%
\nomenclature{$T$}{temperature ($^{\circ}C$ or $K$)}%
\nomenclature{$Q_{g}$}{heat generation ($W/m^{3}$)}
\hspace{-2mm}where $\kappa$ stands for the thermal conductivity, and $Q_{g}$ the volumetric energy generation (or depletion if negative). The Laplacian term in the left-hand side of Eq. \ref{1} physically stands for heat flux gradients. In principle, its value can be equally influenced by three terms, namely, energy storage rate, temperature-dependent thermal conductivity, and energy generation. When the second term and the third terms are manipulated, energy systems are known as those devices aforementioned, respectively. In the literature, however, manipulations of the first term for beneficial applications appear to have been rarely reported.\\
\indent Applications of maximizing or controlling effective thermal conductivities abound. They include areas of micro-electro-mechanical systems (MEMS)\cite{spearing2000materials,ho1998micro}, thermal signals, thermostats, among others. High thermal conductivities for single-walled nanotubes based on MD simulations are reported, promising efficient thermal managements in nanotube-based MEMS devices \cite{che2000thermal}.\\
\indent Another possible application of transient-behavior manipulations is to enhance the effective thermal capacity of the system. In taking advantage of the energy-storage rate in Eq. (\ref{1}), the mass, flipping frequency, heat transfer coefficient, or surface area of the system can be manipulated such that the effective thermal capacity also increases by a few orders of magnitude over the intrinsic material property.
\section{Theoretical concepts}\label{Theoretical concepts}
\subsection{Four-stroke heating and cooling transient-phase cycles}\label{Four-stroke heating and cooling transient-phase cycles}
\begin{figure}
  \centering
  \subfigure{\includegraphics[width=0.5\textwidth]{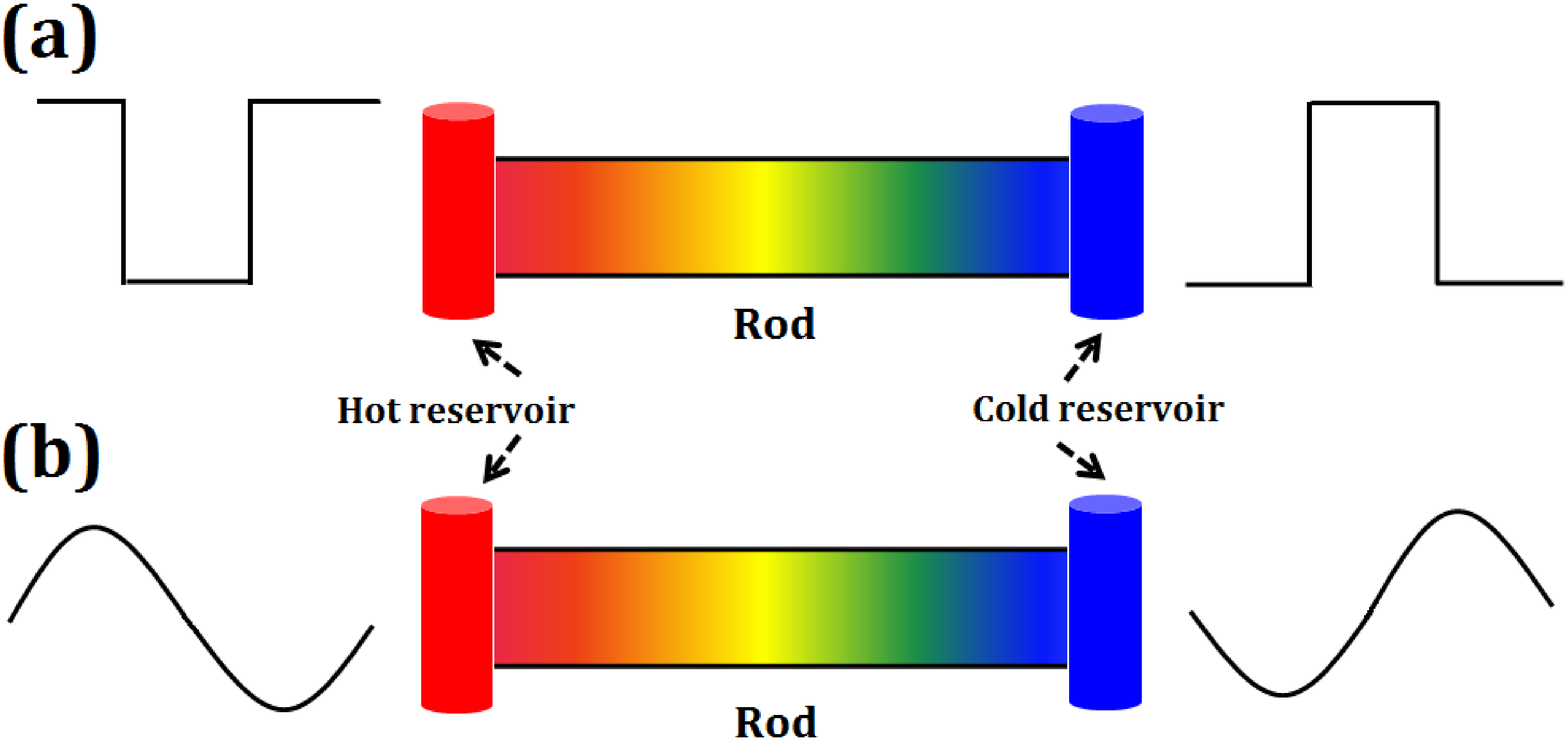}}\\
  \subfigure{\includegraphics[width=0.5\textwidth]{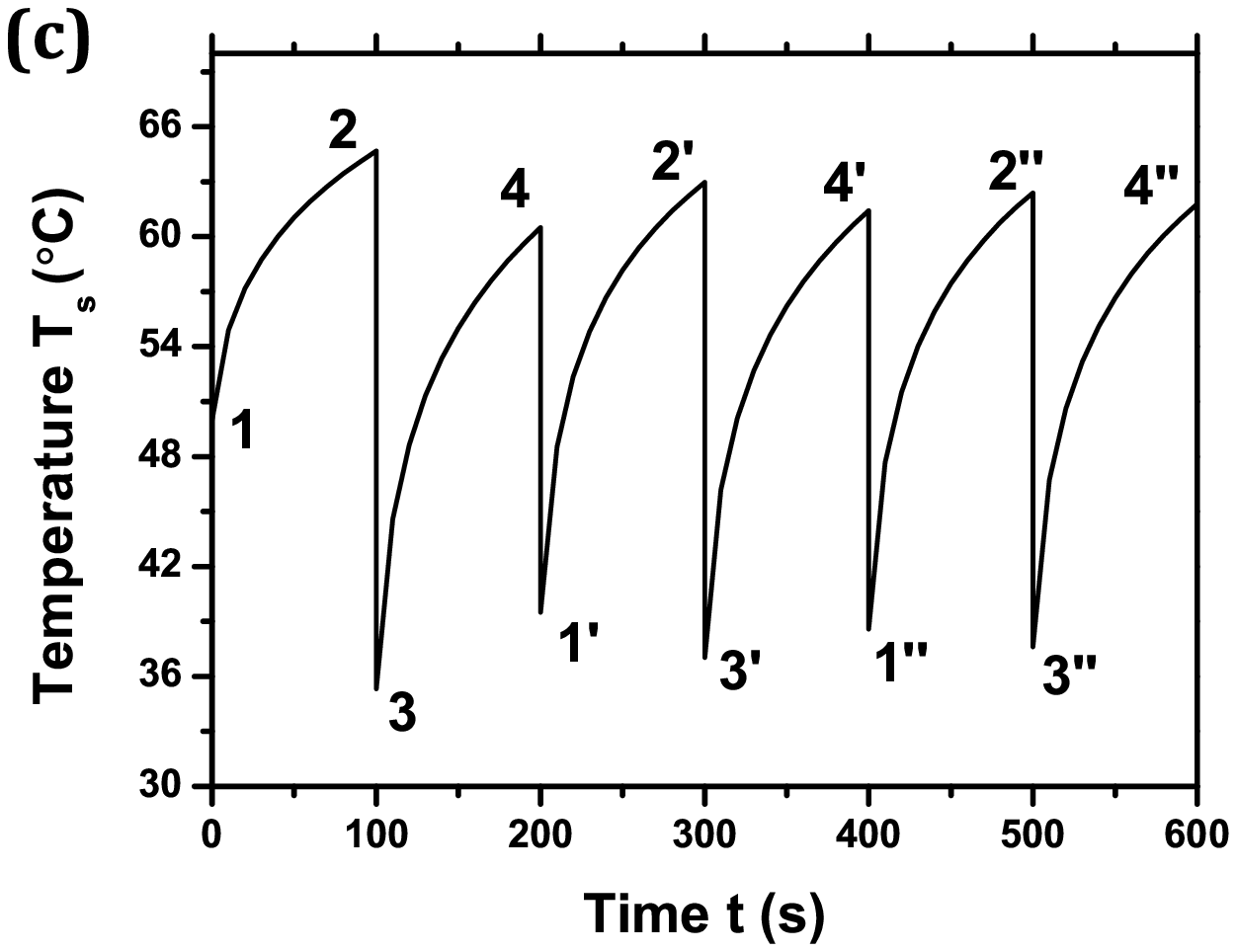}}\\
  \caption{The universe that consists of the rod (or slab) system and two thermal reservoirs. $(a)$ the schematic of the universe and the flipping system; $(b)$ the system remains stationary, but boundary conditions vary continuously; $(c)$ the temperature at the left end of the rod as a function of time.}\label{PIC1}
\end{figure}
Analyses of transient multi-dimensional problems generally require numerical simulations. The description of theoretical concepts, however, is best facilitated by considering transient $1D$ heat conduction phenomena within a rod system insulated circumferentially, sandwiched, and flipped between two thermal reservoirs, as shown in Fig. \ref{PIC1}(a). Mathematically, flipping the rod system while maintaining reservoir temperatures unaltered is similar to keeping the rod stationary while altering reservoir temperatures. Both step-varying and continuously-varying boundary conditions are considered. In the steady state, the heat transfer rate can be readily obtained as
\begin{equation}\label{2}
J=A_{c}\left(T_{Rh}-T_{Rc}\right)/R,
\end{equation}
\nomenclature{$J$}{heat transfer (or heat flow) rate ($W$)}
\nomenclature[S]{$Rh$}{hot reservoir}
\nomenclature[S]{$Rc$}{cold reservoir}
\nomenclature{$A_{c}$}{cross-sectional area ($m^{2}$)}
\nomenclature{$R$}{overall thermal resistance ($m^{2}\,K\,W^{-1}$)}
\nomenclature{$f$}{frequency ($s^{-1}$)}
\hspace{-5mm} where $T_{Rh}$ denotes the temperature of the hot reservoir on the left; $T_{Rc}$ the temperature of the cold reservoir on the right; $A_{c}$ the cross-sectional area of the system; and $R$ the overall resistance, derived to be equal to $1/h_{h}+L/\kappa+1/h_{c}$, where $h$ is commonly known as the heat transfer coefficient such that, at the left end,\\
\begin{equation}\label{3}
h\left(T_{Rh}-T_{s}(t)\right)=-\kappa\left(dT/dx\right)_{x=0},
\end{equation}
\nomenclature[S]{$s$}{at the left surface of the rod system}
\hspace{-1mm}with $T_{s}$ being the temperature on the left end of the system. A similar condition applies to the right end. Furthermore, when a system is steadfastly operating between two reservoirs, its temperature will vary between $T_{Rh}$ and $T_{Rc}$. Under the constraint that none of the pertaining parameter values, including $A_{c}$, $T_{Rh}$, $T_{Rc}$, $h$, $L$, and $\kappa$, is allowed to change, two challenges are sought: $(a)$ to increase the effective thermal conductivity such that $J$ can be increased, and $(b)$ to increase the effective heat capacity such that the operating temperature range can be narrowed. Answers to these two challenges seem to both point to the possibility of manipulating transient behaviors of the rod system, as suggested by Eq. (\ref{1}), and are the crux of the present analysis.\\
\indent A few conditions will be idealized without sacrificing the essential physics:\\
$(a)$ The time required to flip the rod system is negligible.\\
$(b)$ The process of flipping is adiabatic.\\
$(c)$ Values of heat transfer coefficients at $x=0$ and $x=L$ are given the same, rendering the temperature distribution anti-symmetrical. Therefore, at $x=L/2$, the temperature is simply $(T_{Rh} + T_{Rc})/2$, and our attention needs to be paid to only the left half of the system. Examples with $h_{h}\neq h_{c}$ will be considered only when numerical simulations are conducted, otherwise the essential physics may be overwhelmed by nonessential complicities.\\
\indent The cycle of the present energy system consists of four distinctive strokes. For purposes of easy understanding, a practical example is presented, with data pertaining to a Polymethylmethacrylate (PMMA) rod system, ambient conditions, numerical time steps, and the numerical grid given in Table \ref{table1}. All descriptions below are referred to the left end of the rod only, not the entire rod.
\begin{table*}
\begin{center}
\begin{tabular}{cccccc}
\hline
PMMA &thermal conductivity &density &heat capacity &length &diameter\\
rod system &$\kappa\,(W/mK)$ &$\rho\,(kg/m^3$) &$c_{v}\,(J/kgK)$ &$L\,(m)$ &$d\,(m)$\\
 &0.192 &1180 &1450 &0.02 &0.005\\\hline
 &temperature of left &temperature of right &heat transfer coefficient &radiation &\\
ambient condition&hot reservoir $T_{Rh}$ $(^{\circ} C)$ &cold reservoir $T_{Rc}$ $(^{\circ} C)$ &$h\,(W/(m^{2}K))$ &neglected &\\
 &100 &0 &20 & &\\\hline
 &flipping frequency &flipping period &computational time step & &\\
time &$f\,(s^{-1})$ &$t_{o}\,(s)$ &$\Delta t\,(s)$ & &\\
 &0.01 &100 &10 & &\\\hline
numerical &nx &$\Delta x\,(=L/nx)$ &grid staggeredness & &\\
grid &60 &$3.33\times10^{-4}$ &$T(1)$ at $x=0$; $J(1)$ at $x=\Delta x/2$ & &\\\hline
\end{tabular}
\caption{Relevant data for the problem of PMMA rod-system flipping.}
\label{table1}
\end{center}
\end{table*}\\
State $1$: Initially, in reference to Fig. \ref{PIC1}(c), the system is at $50\,^{\circ}C$ uniformly. Suddenly, it is brought in contact with two thermal reservoirs.\\
Process $1-2$: The rod undergoes hot-leg reservoir heating, since the heating is caused by the heat transfer from the hot reservoir and its temperature remains mostly higher than $50\,^{\circ}C$.\\
State $2$: The left end is heated to $64.7\,^{\circ}C$.\\
Process $2-3$: The rod experiences drastic flipping cooling caused by the flipping motion.\\
State $3$: The left end is cooled to $35.3\,^{\circ}C$ $(i. e.,\,100\,^{\circ}C-\,64.7\,^{\circ}C)$.\\
Process $3-4$: This stroke will be recognized as cold-leg reservoir heating, because $T_{s}$ remains mostly colder than $50\,^{\circ}C$.\\
State $4$: The temperature, $T_{s}$, has reached $62.0\,^{\circ}C$. In addition, with certainty, the left end of the system will reach a temperature higher than $50\,^{\circ}C$ at state $4$, because during the hot-leg reservoir heating process, the system has been heated up by $14.7\,^{\circ}C$. Now the system is subject to larger temperature differences between the hot reservoir and the left end. Clearly, the energy that flows into the system during $3-4$ should be larger than the counterpart during $1-2$. In short, $(62.0-35.3)>(64.7-50)$.\\
Process $4-1'$: The system undergoes mild flipping cooling, because the temperature of the system decreases relatively mildly by the flipping mechanism.\\
\subsection{Two-stroke quasi-steady state}\label{Quasi-steady state}
State $1'\rightarrow$ State $2'\rightarrow$ State $3'\rightarrow$ State $4'\rightarrow$ State $1''\rightarrow$$\cdots$$\rightarrow$ State $4^{n}$: The system basically repeats the cycle, except that, gradually, the drastic flipping cooling process will become increasingly mild, and the mild flipping cooling process will oppositely become drastic. When these two processes have merged, the system reaches a quasi-steady state, with $T_{s}$ oscillating between two final temperatures. The valley and the peak will be designated as $T_{\alpha}$ and $T_{\beta}$, respectively. In fact, generally, an energy system that is alternatingly immersed in two (or more) thermal reservoirs will behave similarly. In this case, $T_{\alpha}=62.01\,^{\circ}C$ and $T_{\beta}=37.99\,^{\circ}C$. A four-stroke machine becomes a two-stroke one.\\
\subsection{Entropy of the universe}\label{Entropy of the universe}
Whenever cycles of energy systems are studied, it is constructive to examine if all processes do satisfy the second law of thermodynamics, especially when transient states prevail. The universe of the investigated thermodynamic system can be assumed to consist of two thermal reservoirs and the rod system itself. Hence, the temporal infinitesimal entropy change of the universe can be written as
\begin{equation}\label{4}
dS_{univ}=dS_{Rh}+dS_{Rc}+dS_{sys},
\end{equation}
\nomenclature{$S$}{entropy ($J\,K^{-1}$)}
\nomenclature[S]{$univ$}{universe (reservoirs $+$ the system)}
\nomenclature[S]{$sys$}{system}
\hspace{-3mm}where $dS$ denotes the infinitesimal entropy change; the subscript ``univ'' ``universe''; and ``sys'' ``system''. For the intention of obtaining analytical solutions, the length of the rod is taken to be only $10^{-4} m$, such that the temperature of the rod can be assumed uniform. Furthermore, due to the temperature uniformity, whether or not the rod system is flipped no longer matters. The cross-sectional area is assumed to be $1 m^{2}$ so that it vanishes in equations. When the length of the rod (or the thickness of the disk) increases, and the temperature no longer remains uniform, obtaining analytical solutions becomes increasingly difficult. The fundamental concept, however, should remain the same. For internally reversible processes, the second law of thermodynamics states
\begin{equation}\label{5}
dS=\delta Q/T,
\end{equation}
\nomenclature{$Q$}{heat transfer (or energy) ($J$ or $kJ$)}
\hspace{-1mm}where $\delta Q$ is the net heat transfer into a given control volume. Adopting the symbol ``$\delta$'' instead of ``$d$'' indicates the fact that the value of the heat transfer between two arbitrary states depends on the path, whereas the entropy change does not, and only depends on the two end states.\\
\indent Therefore, heat transfer exiting the hot reservoirs, heat transfer entering the cold reservoir, and the net heat transfer entering the system should be, respectively,
\begin{equation}\label{6}
\begin{split}
&\delta Q_{Rh}=h_{h}\left(T_{Rh}-T_{s}\right)dt;\,\delta Q_{Rc}=h_{c}\left(T_{s}-T_{Rc}\right)dt;\\
\text{and } &\delta Q_{s}=\delta Q_{Rh}-\delta Q_{Rc},
\end{split}
\end{equation}
\nomenclature{$h$}{heat transfer coefficient ($W\,m^{-2}\,K^{-1}$)}
\hspace{-1mm}where signs must be handled carefully. For example, the energy entering the hot reservoir should be $-\delta Q_{Rh}$. Substituting Eq. (\ref{6}) individually first into Eqs. (\ref{5}) then into Eq. (\ref{4}) yields
\begin{equation}\label{7}
\frac{dS_{univ}}{dt}=-d_{1}+d_{2}T(t)+\frac{d_{3}}{T(t)},
\end{equation}
which can be integrated to become
\begin{equation}\label{8}
S_{univ}=S_{o}+\left(d_{2}d_{5}-d_{1}\right)t+\frac{d_{2}}{d_{4}}\left(T_{o}-T(t)\right)+\frac{d_{3}}{d_{4}d_{5}}ln\frac{T(t)}{T_{o}\exp(-d_{4}t)}, \end{equation}
\nomenclature{$d_{1}, \cdots, d_{5}$}{coefficients that appear in Eq. (\ref{8}) (various dimensions)}
\begin{table}
\begin{center}
\begin{tabular}{cccc}
\hline
$d_{1}$ &$2(h_{h}+h_{c})$ &$d_{2}$ &$h_{h}/T_{Rh}+h_{c}/T_{Rc}$\\\hline
$d_{3}$ &$h_{h}T_{Rh}+h_{c}T_{Rc}$ &$d_{4}$ &$(h_{h}+h_{c})/(mcv)$\\\hline
$d_{5}$ &$(h_{h}T_{Rh}+h_{c}T_{Rc})/(h_{h}+h_{c})$ & &\\\hline
\end{tabular}
\caption{Expressions of $d_{1}$ to $d_{5}$ contained in Eqs. (\ref{7}) and (\ref{8}).}
\label{table2}
\end{center}
\end{table}
\hspace{-1mm}where $T(t)= (T_{o}-d_{5})\exp(-d_{4}t)+d_{5}$; $T_{o}$ and $S_{o}$ are reference values; $d_{1},\,\cdots,\,d_{5}$ are listed in Table \ref{table2}. When $t= t_{ss}$, the rod system has reached the steady state, implying that its entropy does not vary any more. The energy transferred from the hot reservoir to the system can be simply written as $Q_{Rh}= h_{h}(T_{Rh}-T_{ss})t=Q_{Rc}$, leading to
\begin{equation}\label{9}
\Delta S_{univ,ss}=h_{h}\left(T_{Rh}-T_{ss}\right)\Delta t\left(1/T_{Rc}-1/T_{Rh}\right),
\end{equation}
\nomenclature[S]{$ss$}{steady state}
\hspace{-1mm}which increases linearly with $\Delta t$. It is found that the value of $S_{univ}(t+\Delta t)-S_{univ}(t)$, obtained from Eq. (\ref{8}), is equal to that obtained from Eq. (\ref{9}), as long as $t>t_{ss}$. For example, for $\Delta t=60\,s$ and $t= 2060\,s$, both values are equal to $29.4612\,J/K$. Such identicalness suggests that the present analysis obeys the second law, and that the quasi-steady state will be reached.\\
\begin{figure}
\centering
\includegraphics[width=0.5\textwidth]{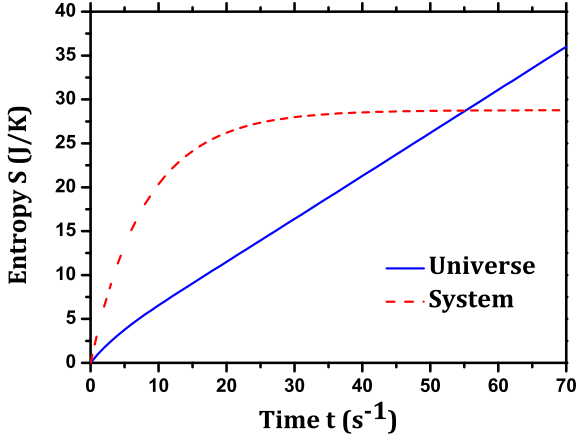}
\caption{The entropy versus time, with relevant data listed in Table \ref{table1}.}\label{PIC2}
\end{figure}
\indent In Fig. \ref{PIC2}, $S_{univ}(t)$, along with the entropy of the system, namely $S_{sys}(t)= S_{sys,o}+ mc_{v}\ln(T(t)/T_{i})$, are plotted versus $t$. Both reference values, $S_{univ,o}$ and $S_{sys,o}$ are taken to be zero for convenience. As $t$ increases, $S_{sys}$ gradually levels off. After the steady state has reached, it becomes constant. However, the entropy of the universe continues to increase, because there will be permanently a constant amount of energy transferring from the hot reservoir via the rod (or the disk) to the cold reservoir. As long as there is heat transfer taking place within the universe, $S_{univ}$ will increase.
\subsection{Augmentation ratios}\label{Augmentation ratios}
\begin{figure}[b]
\centering
\includegraphics[width=0.5\textwidth]{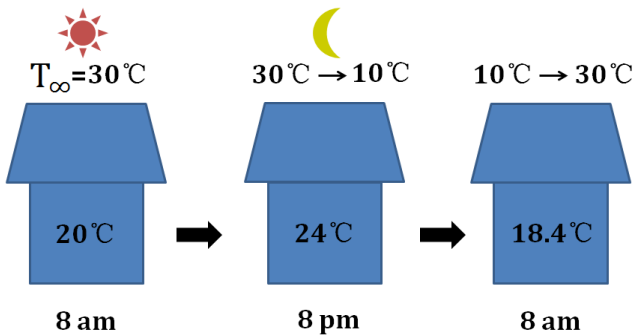}\\
\caption{Heating and cooling of the house subject to oscillatory ambient conditions for the purpose of defining and explaining the nonlinear thermal bias (NTB).}\label{PIC3}
\end{figure}
At this juncture, it is conducive to introduce, define, and explain a term named ``nonlinear thermal bias'', which plays an essential role in commonly-encountered phenomena when an energy system is immersed in more than one thermal reservoir alternatingly ($n\geq2$). Such phenomena can be typified by heating a house ($40\,m\times40\,m\times5\,m$) during the daytime ($T_{outdoors}=30\,^{\circ}C$) and cooling it during the nighttime ($T_{outdoors}=10\,^{\circ}C$), as schematically shown in Fig. \ref{PIC3}. Relevant data include $c_{v}=1000\,J/(kgK)$ and $h=10W/(m^{2}K)$ with the floor insulated. Among them, a property that can be conveniently manipulated is the density of the house. Imagine that rocks are intentionally stored inside the house such that the true density of the house becomes $845.7\,kg/m^{3}$ (instead of $1\,kg/m^{3}$ for air). Initially, the house is maintained at $20\,^{\circ}C$. Suddenly, it is immersed in an outdoor airflow at $30\,^{\circ}C$. After it is heated up for $12$ hours, the outdoor temperature drops to $10\,^{\circ}C$ suddenly, and remains the same for another $12$ hours. The daily cycle repeats. The following string of numbers represents the history of temperature variations:
\begin{equation}
\begin{split}
&20.000^{\circ}C\rightarrow(\text{heating for $12$ hour})\,24.000\rightarrow(\text{cooling for}\,12\\
&\text{hours})\,18.400\,\rightarrow\,23.040\,\rightarrow\,17.824\,\rightarrow\,22.694\,\rightarrow\,\cdots\,\rightarrow\\
&17.500\rightarrow\,22.500\rightarrow \text{oscillating between}\,17.5\,^{\circ}C\,\text{and}\,22.5\,^{\circ}C\\
&\text{quasi-steadily}. \nonumber
\end{split}
\end{equation}
\indent The former of the two final values can be obtained analytically as
\begin{equation}\label{10}
T_{\alpha}=T_{Rh}-\Delta Texp\left(-c_{1}t_{qs}\right),
\end{equation}
\nomenclature[S]{$\alpha$}{the valley temperature in the quasi-steady state}
\nomenclature[S]{$qs$}{quasi-steady state}
\nomenclature{$c_{1}$}{a convenient parameter defined as $hA/(mcv)$ ($s^{-1}$)}
\hspace{-2mm}whereas the latter becomes
\begin{equation}\label{11}
T_{\beta}=T_{Rh}-\Delta Texp\left[-c_{1}\left(t_{qs}+t_{o}\right)\right],
\end{equation}
\nomenclature[S]{$\beta$}{the peak temperature in the quasi-steady state}
\nomenclature{$t_{o}$}{the flipping period ($s$)}
\hspace{-1.5mm}where $\Delta T=T_{Rh}-T_{i}$; $T_{i}$ is the initial temperature of the house; $t_{qs}$ the time for the house to start entering the quasi-steady phase; and $c_{1}=hA/(mc_{v})$. Setting $t_{qs}$ to zero, $T_{\alpha}+T_{\beta}$ to $T_{Rh}+T_{Rc}$, and $T_{i}$ to $T_{\alpha}$ leads to (see \ref{Derivation of Eq.(12)})
\begin{equation}\label{12}
T_{\alpha}=T_{Rh}-\frac{T_{Rh}-T_{Rc}}{1+exp\left(-c_{1}t_{o}\right)}.
\end{equation}
\indent The capacity augmentation ratio is defined as
\begin{equation}\label{13}
r_{cap}=\frac{c_{v,eff}}{c_{v}}=\frac{T_{Rh}-T_{Rc}}{T_{\beta}-T_{\alpha}},
\end{equation}
\nomenclature[S]{$cap$}{heat capacity}
\nomenclature[S]{$eff$}{effective}
\hspace{-1mm}because $mc_{v,eff}(T_{\beta}-T_{\alpha})=mc_{v}(T_{Rh}-T_{Rc})$. It can also be used to estimate the thermal fatigue and the energy savings. For example, consider a piston-cylinder system operating between a hot reservoir at $100\,^{\circ}C$ and a cold one at $0\,^{\circ}C$ alternatingly. It may be unnecessary to estimate the thermal-fatigue tolerance to be a temperature interval of $[0\,^{\circ}C, 100\,^{\circ}C]$, because the bulk of the system will remain within the interval of $[T_{\alpha},T_{\beta}]$.\\
\indent When the temperature in the system is non-uniform, such as in the case of the PMMA rod, the conductivity augmentation ratio is defined as
\begin{equation}\label{14}
r_{cond}=\frac{\kappa_{eff}}{\kappa}=\frac{J_{qs}}{J_{ss}}=\frac{\int_{0}^{t_{o}}\left[T_{Rh}-T_{s}(t)\right]dt}{(T_{Rh}-T_{ss})t_{o}}.
\end{equation}
\nomenclature[S]{$cond$}{thermal conductivity}
\indent The steady-state temperature at $x=0$, $T_{ss}$, can be derived by taking the energy balance at the interface $x=0$ as $h(T_{Rh}-T_{ss})=2\kappa[T_{ss}-0.5(T_{Rh}+T_{Rc})]/L$, leading to
\begin{equation}\label{15}
T_{ss}=\frac{T_{Rh}\left(B_{i}+1\right)+T_{Rc}}{B_{i}+2},
\end{equation}
\nomenclature{$B_{i}$}{Biot number defined as $hL/\kappa$ (dimensionless)}
\hspace{-1mm}where $B_{i}$ is known as the Biot number, defined as $B_{i}=hL/\kappa$. When $\delta=0$, it can be shown that (see \ref{Derivation of Eq.(16)}),
\begin{equation}\label{16}
r_{cond}=\frac{B_{i}}{2}+1.
\end{equation}
\subsection{Nonlinear thermal bias}\label{Nonlinear thermal bias}
Finally, once $T_{\alpha}$ and $T_{\beta}$ are determined for a universe, the mean temperature during the reservoir heating (or cooling) process can be readily evaluated by
\begin{equation}\label{17}
Q=A_{c}h\int_{0}^{t_{o}}\left(T_{Rh}-T(t)\right)dt=Ah\left(T_{Rh}-T_{\alpha}\right)\left(1-\exp(-c_{1}t_{o})\right)/c_{1},
\end{equation}
\nomenclature{$A$}{area of the house $(m^{2})$}
\hspace{-1mm}where the analytical exponential expression is valid only for uniform-temperature cases, e.g., the house heating/cooling. It is worth noting that the time-averaged temperature during the heating process is by no means equal to $(T_{Rh} + T_{Rc})/2$, because $T(t)$ is a nonlinear function of time. This quantity will be termed as nonlinear thermal bias (NTB), and is defined as
\begin{equation}\label{18}
NTB=\int_{0}^{t_{o}}T(t)dt/t_{o}.
\end{equation}
\nomenclature{$NTB$}{nonlinear thermal bias ($K$)}
\indent Once it is determined, it can be used to conveniently calculate the heat transfer during the quasi-steady heating process as
\begin{equation}\label{19}
Q=A_{c}h\left(T_{Rh}-NTB\right)t_{o}.
\end{equation}
\indent For the problem of house heating, $NTB = 20.2119\,^{\circ}C$, and $Q= 3.3828\times10^{7}\,kJ$, which happens to be equal to $\Delta U=mc_{v}(T_{\beta}-T_{\alpha})$ as well according to the energy conservation. By symmetry with respect to $20\,^{\circ}C$, during the cooling process, $NTB = 19.7881\,^{\circ}C$ and $Q$ bears the same value, but with the negative sign. In addition, $NTB$ can be used for purposes of saving energy and feeling comfortable. Suppose that the owner of the house wishes to set the thermostat at $23.5\,^{\circ}C$ during a heating process. Then he may manipulate the value of $c_{1}$ such that $NTB$ is equal to $23.5\,^{\circ}C$, because, with this manipulation, the temperature of this house will linger around $23.5\,^{\circ}C$ most frequently.
\subsection{Buoyancy-driven recirculating flows when $T_{top}>T_{bottom}$}
\label{Buoyancy-driven recirculating flows when $T_{top}>T_{bottom}$}
In a typical $3D$ enclosure containing a fluid, the buoyancy force can hardly be induced if the top face is hot and the bottom face is cold. The ocean with its surface at $23\,^{\circ}C$ and its bottom at $3\,^{\circ}C$ appears to simulate such an enclosure. However, the temperature gradient in the ocean does exist, and should be legitimately regarded as an energy potential waiting to be harnessed. Furthermore, the second law of thermodynamics does not dictate that it be impossible to build a $2T$ engine when $T_{top}> T_{bottom}$. Hence, a preliminary feasibility study is conducted herein to face this challenge. It explores the possibility of utilizing transient behaviors of a fictitious propeller-like machine situated in the enclosure subject to the condition of cold bottom and hot top. It requires solving a set of partial differential equations that govern mass conservation, momentum transports in $3$ Cartesian directions, energy conservation, and the density-state equation written as \cite{tao2001numerical}
\begin{equation}\label{20}
\frac{\partial \rho}{\partial t}+\nabla\cdot\left(\rho\bm{\bar{\upsilon}}\right)=0,
\end{equation}
\nomenclature{$\bm{\bar{\upsilon}}$}{flow velocity, $ui+vj+wk$ ($m\,s^{-1}$)}
\begin{equation}\label{21}
\frac{\partial}{\partial t}\left(\rho\bm{\bar{\upsilon}}\right)+\rho\bm{\bar{\upsilon}}\cdot\left(\nabla\bm{\bar{\upsilon}}\right)=\mu\nabla^{2}\bm{\bar{\upsilon}}-\nabla p+\rho \bm{\bar{g}},(\text{for}\,\,\vec{i},\,\vec{j},\,\vec{k})
\end{equation}
\nomenclature[G]{$\mu$}{viscosity ($kg\,m^{-1}\,s^{-1}$)}
\nomenclature{$p$}{pressure ($N\,m^{-2}$)}
\nomenclature{$\bm{\bar{g}}$}{gravitational acceleration $(m\,s^{-2})$}
\begin{equation}\label{22}
\frac{\partial}{\partial t}\left(\rho c_{v}T\right)+\rho c_{p}\bm{\bar{\upsilon}}\cdot\left(\nabla T\right)=\kappa\nabla^{2}T,
\end{equation}
\nomenclature{$c_{p}$}{heat capacity with pressure kept constant\,($Jkg^{-1}K^{-1}$)}
and\\
\begin{equation}\label{23}
\rho=1000.5262-0.1390T\,\left(T\,\text{in C}\right),
\end{equation}
where $\bar{g}=-g\vec{j}$. Results demonstrate that a small amount of thermal energy released at the bottom of enclosure is sufficient to allow the buoyancy force to overcome the viscous force, and thus to possibly sustain the system flipping or rotating.
\section{Numerical simulations}
\label{Numerical simulations}
\subsection{Transient $3D$ heat conduction subject to system flipping}\label{Transient $3D$ heat conduction subject to system flipping}
\begin{figure}
  \centering
  \includegraphics[width=0.5\textwidth]{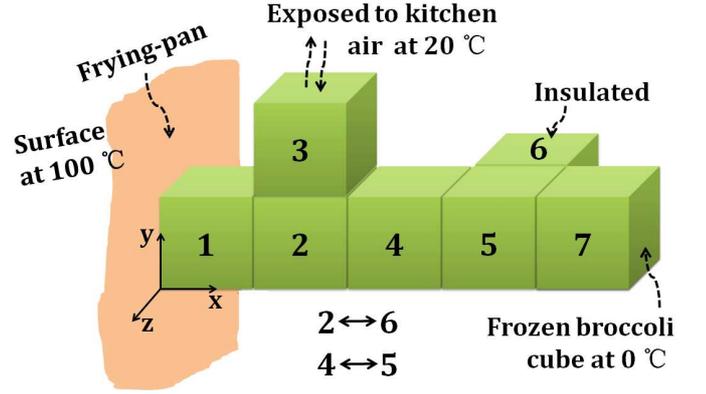}\\
  \caption{System schematic of transient $3D$ heat conduction with the system flipping.}\label{PIC4}
\end{figure}
Frying broccoli chunks in a pan rightfully constitutes a $3D$ transient problem subject to the system flipping. It can be modeled without losing essential physical concepts by considering several individual cubes with cubes $2$ and $6$ being switched and $4$ and $5$ switched during the frying process, as schematically shown in Fig. \ref{PIC4}. Thermal properties are taken to be the same as those of PMMA for convenience. Initially, all cubes are frozen at $0\,^{\circ}C$. Suddenly, cube $1$ touches the frying pan, is heated to $100\,^{\circ}C$, and is maintained at $100\,^{\circ}C$. Only the top side of cube $3$ is exposed to the kitchen air at $20\,^{\circ}C$; elsewhere external sides of all cubes are assumed to be insulated. The flipping frequency is $0.1\,s^{-1}$ (or $t_{o}= 10\,s$). The set of discretized governing equations written in terms of the matrix form is given in \ref{Derivation of $3D$ transient heat conduction equations}. The global energy balance is checked to enhance the confidence in the validity of the MATLAB code. Finally, both the steady-state temperatures and the quasi-steady temperatures (after approximately $10^{5}\,s$) of $7$ cubes are listed in Table $\ref{table3}$. The conductivity augmentation ratio in $x$ direction is computed to be $r_{cond}= 2.710$. As the number of cubes increases, the $r_{cond}$ value is expected to increase because $T_{2}$ for the steady state will become closer to $100\,^{\circ}C$, while $T_{2}$ for the quasi-steady state will remain low due to cube switching. Hence the common sense has it that stirring the food during the frying process can avoid food burning.
\begin{table*}
\begin{center}
\begin{tabular}{cccccccc}
\hline
             &$T_{1}$ &$T_{2}$ &$T_{3}$ &$T_{4}$ &$T_{5}$  &$T_{6}$ &$T_{7}$ (in $\,^{\circ}C$) \\
steady state &$100$   &$63.73$ &$48.76$ &$42.48$ &$21.25$  &$21.25$ &$0$ \\\hline
quasi-steady(1) &$100$   &$2.291$ &$8.356$ &$1.152$ &$1.139$  &$1.139$ &$0$ \\ \hline
quasi-steady(2) &$100$   &$1.139$ &$8.3556$ &$1.139$ &$1.152$  &$2.291$ &$0$ \\ \hline
\end{tabular}
\caption{Nodal temperature solution of the transient $3D$ heat conduction with two pairs of cube switching.}
\label{table3}
\end{center}
\end{table*}
\subsection{Transient $1D$ heat conduction with flipping systems}\label{Transient $1D$ heat conduction with flipping systems}
When behaviors of the energy system do not largely depend on the dimensionality of the problem, an internally-developed $1D$ transient code has also been used to save computational time. Herein, Eq. (\ref{1}) degenerates into its $1D$ counterpart, and can then be discretized to become
\begin{equation}\label{24}
mc_{v}\left(T_{i}-T_{i}^{p}\right)/\Delta t=kA_{c}\left(T_{i-1}-T_{i}\right)/\Delta x-kA_{c}\left(T_{i}-T_{i+1}\right)/\Delta x,
\end{equation}
$i=2,3,\cdots,nx$\\
\nomenclature[S]{$i$}{$i$th node or $i$th particle, or at the initial state}
\indent subject to the left-side boundary, ($i=1$),
\begin{equation}\label{25}
\frac{1}{2}mc_{v}\left(T_{1}-T_{1}^{p}\right)/\Delta t=hA_{c}\left(T_{Rh}-T_{1}\right)-kA_{c}\left(T_{1}-T_{2}\right)/\Delta x,
\end{equation}
and the right-side boundary, ($i=nx+1$),
\nomenclature{$nx$}{number of grid intervals (dimensionless)}
\begin{equation}\label{26}
\frac{1}{2}mc_{v}\frac{\left(T_{nx+1}-T_{nx+1}^{p}\right)}{\Delta t}=kA_{c}\frac{\left(T_{nx}-T_{nx+1}\right)}{\Delta x}-hA_{c}\left(T_{nx+1}-T_{Rc}\right).
\end{equation}
\indent In Eqs. (\ref{24})-(\ref{26}), the superscript ``$p$'' stands for ``at the previous time step''. Initially, the rod system is maintained at $50\,^{\circ}C$ throughout. After the set of $nx+1$ linear equations are solved by a direct matrix-inversion solver, nodal temperatures $T_{i}^{p}$ ($i=1, 2, 3, \cdots, nx+1$) are immediately updated as $T_{i}$. At the instant of flipping, however, they are updated according to $T_{1}= T_{nx+1}, T_{2}= T_{nx}, T_{3}= T_{nx-1}, \cdots, T_{nx+1}= T_{1}$. Grid nodes for temperatures coincide with the $x$ coordinate, i. e., $T(1)$ is located at $x=0$, whereas nodal conductive fluxes stagger for half grid intervals. Consequently, the factor of $1/2$ appears in Eqs. (\ref{25}, \ref{26}).
\subsection{Hamiltonian oscillators}\label{Hamiltonian oscillators}
In areas of MEMS, ultrafast Laser heating \cite{petrova2006temperature,glezer1997ultrafast,chen2006semiclassical}, non-Fourier thermal transports \cite{chang2008breakdown,cao2007equation,shiomi2006non}, and temperature-dependent thermal properties \cite{das2003temperature,mintsa2009new}, it may be insufficient to analyze macro-scale systems alone to further understand fundamental mechanisms that are related to both the first law and the second law of thermodynamics. Studies of micro-scale systems are called for. They are typified by examining $1D$ heat conduction idealized as a string of Hamiltonian oscillators \cite{lepri2003thermal,dhar2008heat} moving in a lattice. The Hamiltonian is defined as
\begin{equation}\label{27}
H=\sum_{i}^{N}\left[\frac{P_{i}^{2}}{2m}+\frac{\beta}{4}x_{i}^{4}+\frac{k}{2}\left(x_{i+1-x_{i}}\right)^{2}\right],
\end{equation}
\nomenclature{$P_{i}$}{momentum of the $i$th particle ($kg\,m\,s^{-1}$)}
\nomenclature[G]{$\beta$}{strength of the on-site potential ($kg\,m^{-2}\,s^{-2}$)}
\nomenclature{$k$}{spring constant ($kg\,s^{-2}$)}
\nomenclature{$x_{i}$}{the displacement of the $i$th particle ($nm$)}
\hspace{-2mm}where $N$ is the total number of particles; $m$ the mass of particles; $P_{i}$ the momentum of the $i^{th}$ particle; $x_{i}$ the displacement from the equilibrium position; $\beta$ the strength of the on-site potential; and $k$ the spring constant.\\
\indent In present simulations, fixed boundary conditions are used, and the chain is connected to two thermal reservoirs at temperatures $T_{Rh}=2$, and $T_{Rc}=1$, respectively. Langevin thermal reservoirs \cite{hatano2001steady,sekimoto1997kinetic} are used and equations of motion are integrated by using the fourth-order stochastic Runge-Kutta algorithm \cite{21honeycutt1992stochastic,hull1972comparing}. Boundary conditions for oscillators $i=1$ and $i=64$ are prescribed as
\begin{eqnarray}\label{28}
mx_{1}^{''}=k\left(x_{2}-2x_{1}\right)-\beta x_{1}^{3}+\eta_{w}(t)-\lambda_{w}x_{1}^{'}
\end{eqnarray}
\nomenclature[G]{$\eta$}{intermediate variable in Hamiltonian-oscillator formulation ($N$)}
\nomenclature[G]{$\lambda$}{damping factor ($kg\,s^{-1}$)}
and
\begin{eqnarray}\label{29}
mx_{64}^{''}=k\left(x_{63}-2x_{64}\right)-\beta x_{64}^{3}+\eta_{e}(t)-\lambda_{e}x_{64}^{'},
\end{eqnarray}
where
\begin{eqnarray}
\eta_{w}(t)=\sqrt{-4k_{B}T_{Rh}\lambda_{w}ln\left(a_{1}\right)}\cos(2\pi a_{2})\nonumber
\end{eqnarray}
\nomenclature[S]{$h$}{hot}
and
\begin{eqnarray}
\eta_{e}(t)=\sqrt{-4k_{B}T_{Rc}\lambda_{e}ln\left(a_{3}\right)}\cos(2\pi a_{4}). \nonumber
\end{eqnarray}
\nomenclature{$a_{1},\cdots,a_{4}$}{randomly generated numbers (dimensionless)}
\nomenclature[S]{$c$}{cold, or cross-sectional}
\indent Here $a_{1}$, $a_{2}$, $a_{3}$, and $a_{4}$ are randomly-generated numbers between $0$ and $1$; values of $\lambda_{w}$, $\lambda_{e}$ (damping factors), $\beta$ and $\kappa_{B}$ are all taken to be unity. Simulations are performed sufficiently long to allow the system to reach the steady state. Average values of temperatures and heat fluxes in the system are taken. Then the thermal conductivity is computed according to the Fourier's Law. Finally, through iterations, the value of the spring constant $k$ is adjusted such that this computed thermal conductivity eventually coincides with the macro-scale value.
\begin{figure}[h]
  \centering
  \includegraphics[width=0.5\textwidth]{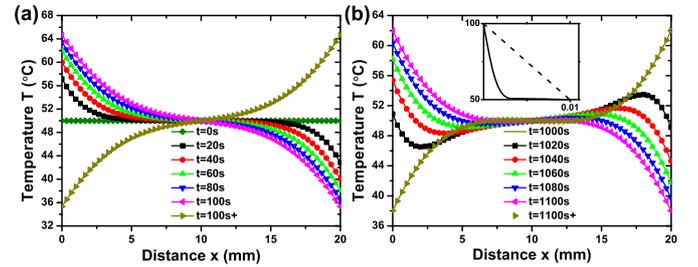}\\
  \caption{Temperature distributions parameterized in time for the flipping rod system. $(a)$ during the first stroke of hot-leg reservoir heating; $(b)$ in the quasi-steady state.}\label{PIC5}
\end{figure}
\subsection{Transient $3D$ buoyancy-driven flows}\label{Transient $3D$ buoyancy-driven flows}
Transient $3D$ buoyancy-driven recirculating flows are governed by Eqs. (\ref{20}-\ref{23}). One of primary challenges to computationally solve these nonlinear partial differential equations is to adequately handle convective terms in transport equations. These terms contain three field variables, namely, the density, the flow velocity, and the temperature (or again the flow velocity), leading to high nonlinearity, especially when large temperature gradients (thus large density gradients) exist. Since the emphasis of the present study is on the thermal physics, but not numerical analyses, a reliable finite-element-based commercial software package, known as COMSOL \cite{comsol2005comsol}, has been used. The $3D$ grid is numerically generated using a companion code. The algorithm adopts the Galerkin formulation, in which weighting functions are taken to be identical to basis functions, and nodal residuals are made orthogonal to these basis functions \cite{shih1984numerical}. The temperature non-uniformity originates from an artificial heat source located at the lower left corner of the enclosure, instead of from the conventional boundary conditions.
\section{Results and discussion} \label{Results and discussion}
\subsection{Maximizing and controlling effective $\kappa$ and $c_{v}$}\label{Maximizing and controlling effective k and cv}
\begin{figure}
  \centering
  \includegraphics[width=0.5\textwidth]{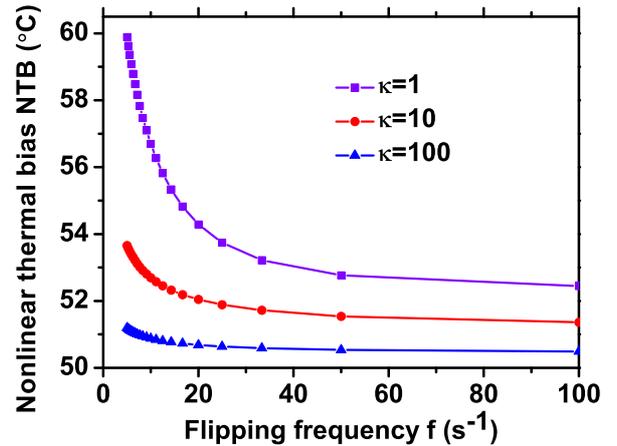}\\
  \caption{The nonlinear thermal bias (NTB) versus the flipping frequency parameterized in $\kappa$.}\label{PIC6}
\end{figure}

\begin{figure}
  \centering
  \subfigure{\includegraphics[width=0.5\textwidth]{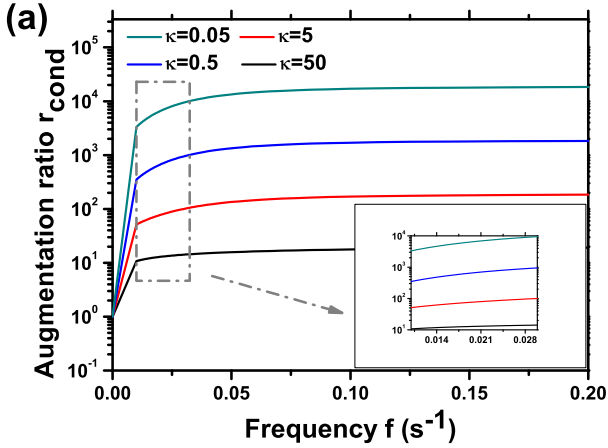}}\\
  \subfigure{\includegraphics[width=0.5\textwidth]{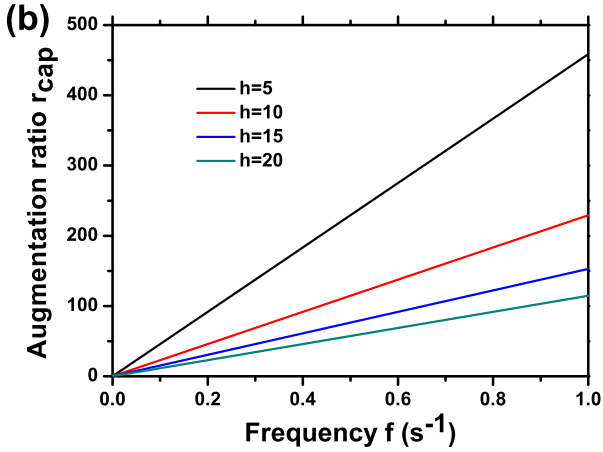}}\\
  \caption{Augmentation ratios versus the flipping frequency. $(a)$ thermal conductivity augmentation ratio versus $f$ parameterized in $\kappa$; (b) heat capacity augmentation ratio versus $f$ parameterized in $h$.}\label{PIC7}
\end{figure}

\begin{figure}
  \centering
  \includegraphics[width=0.5\textwidth]{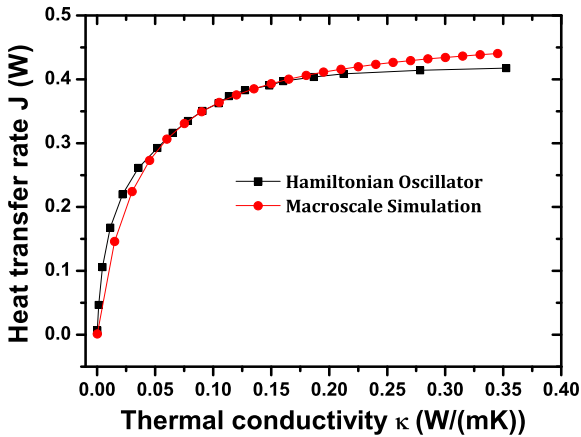}\\
  \caption{The heat transfer rate versus $\kappa$ for both macro-scale systems and Hamiltonian oscillators.}\label{PIC8}
\end{figure}

\begin{figure}
  \centering
  \includegraphics[width=0.5\textwidth]{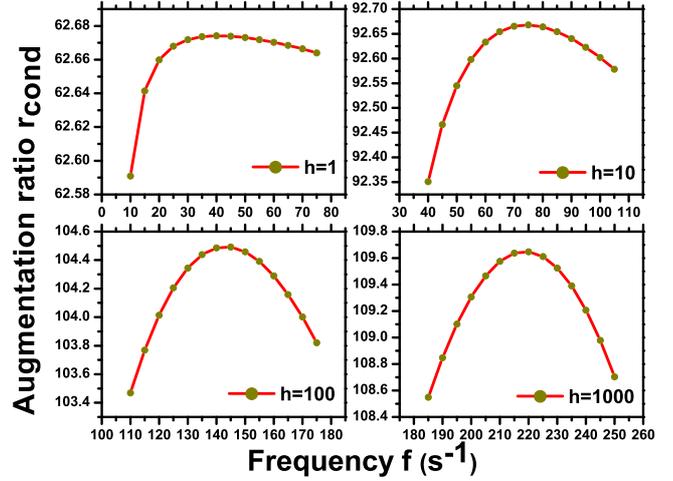}\\
  \caption{Thermal conductivity augmentation ratio versus the boundary-condition oscillation frequency parameterized in the heat transfer coefficient.}\label{PIC9}
\end{figure}

Figure $\ref{PIC5}$ shows the temperature of the rod system versus $x$ parameterized in $t$. In Fig. $\ref{PIC5}(a)$, six profiles at first six time steps are shown during the hot-leg reservoir heating with $\Delta t=20\,s$. After these six time steps, the rod system is flipped, indicating that the flipping period, $t_{o}$, is $100\,s$, or that the flipping frequency, $f$, is $0.01\,s^{-1}$. Figure $\ref{PIC5}(b)$ shows quasi-steady profiles, with $T_{\alpha}=37.99\,^{\circ}C$ and $T_{\beta}=62.01\,^{\circ}C$. It can be observed that the temperature rise in the first time step is much larger than that in the second time step, evidencing the existence of $NTB$. When $h$ approaches infinity, $T_{s}$ will become $T_{Rh}$. The left end of the rod will remain at $T_{Rh}$ almost all the time because it will instantaneously be heated up from $0\,^{\circ}C$ to $100\,^{\circ}C$ after the flipping due to the existence of an infinite $h$ value. The adjacent nodal temperature, $T(2)$, however, will hover in the neighborhood of $50\,^{\circ}C$, thus establishing an extremely steep temperature gradient near $x=0$, as shown in the inset of Fig. \ref{PIC5}$(b)$. In reference to the steady-state solution represented by the dashed line, the energy balance at $x=0$ can be written as $h(100-T_{s})=k(T_{s}-51)/\Delta x$. If $T_{s}=99\,^{\circ}C$, $h=10000\,W/(m^{2}K)$, $L=1\,m$, $\kappa=0.2083\,W/(mK)$, and $\Delta x=0.001$, the thermal conductivity augmentation ratio can be estimated to be nearly $0.5L/\Delta x=500$. This value demonstrates that $r_{cond}$ value can increase by a few orders of magnitude when both values of the thermal resistance $L/\kappa$ and $h$ are large.\\
\indent In Fig.\,$\ref{PIC6}$, NTB is plotted versus the flipping frequency parameterized in $\kappa$. Take $\kappa=1$ and $f=10\,s^{-1}$ for example. The figure indicates that $NTB=56.70\,^{\circ}C$. Hence during the hot-leg heating process, the heat transfer from the hot reservoir to the rod system can be calculated to be $h(T_{Rh}-NTB)t_{o} = 8660\,J$.\\
\hspace{-2.2mm}\indent Figure $\ref{PIC7}$ presents augmentation ratios, $r_{cond}$ and $r_{cap}$, versus the flipping frequency, and constitutes the most important result among all. When $f$ diminishes to zero, the system does not flip. Therefore, all ratios converge to the value of unity, as expected. In Fig. $\ref{PIC7}(a)$, they are parameterized in $\kappa$. As the material becomes less conducting, the benefit of flipping becomes more pronounced. This trend can be readily understood in analogy to cooking in the kitchen. When the soup in the pot becomes more viscous (thus leading to less circulation and lower effective $\kappa$), stirring the soup becomes more necessary to avoid the soup burning. Potentially, for $\kappa=0.05$ (as a comparison, $\kappa_{air}=0.026$), the ratio can reach $10000$. In the inset figure, clarity in variations of $r_{cond}$ is observed. Sometimes, in MEMS, the objective of the design may be adjusting $\kappa$ values, but not necessarily maximizing $\kappa$ values. If so, the flipping frequency can be controlled to achieve such a purpose. In Fig. $\ref{PIC7}(b)$, the result is parameterized in the heat transfer coefficient. As $h$ decreases, again the benefit of controlling reservoir-temperature-oscillating frequencies becomes more pronounced.\\
Figure $\ref{PIC8}$ shows the comparison between the heat conduction rate obtained by Hamiltonian-oscillator simulations and the present numerical simulation (Eqs. $\ref{24}-\ref{26}$). The agreement is fair. The empirical value of the spring constant for the Hamiltonian oscillator is iteratively determined such that the computed macro-scale thermal conductivity is equal to $0.1$.\\
It is noted that results in Fig. $\ref{PIC7}(a)$ are obtained during the quasi-steady phase, with no peaks observed to exist. In other words, both $r_{cond}$ and $r_{cap}$ are monotonic functions of $f$. A natural question may arise: will peaks exist during the transient phase or under continuously-changing (or step-changing) boundary condition? (See \ref{Analytical solution $T(t)$ when $nx=1$ subject to continuously-changing boundary conditions}) Figure $\ref{PIC9}$ shows $r_{cond}$ versus $f$ parameterized in $h$. These results are obtained analytically by taking $nx$ to be unity (i.e., zero interior node). Indeed, peaks are observed, suggesting that, for example, $f$ can be manipulated to be $145\,s^{-1}$ in order to maximize $r_{cond}$ for $h=100\,W/(m^{2}K)$.\\
\subsection{Oceanic buoyancy-driven hydraulic energy}\label{Oceanic buoyancy-driven hydraulic energy}
\begin{figure}[t]
  \centering
  \includegraphics[width=0.45\textwidth]{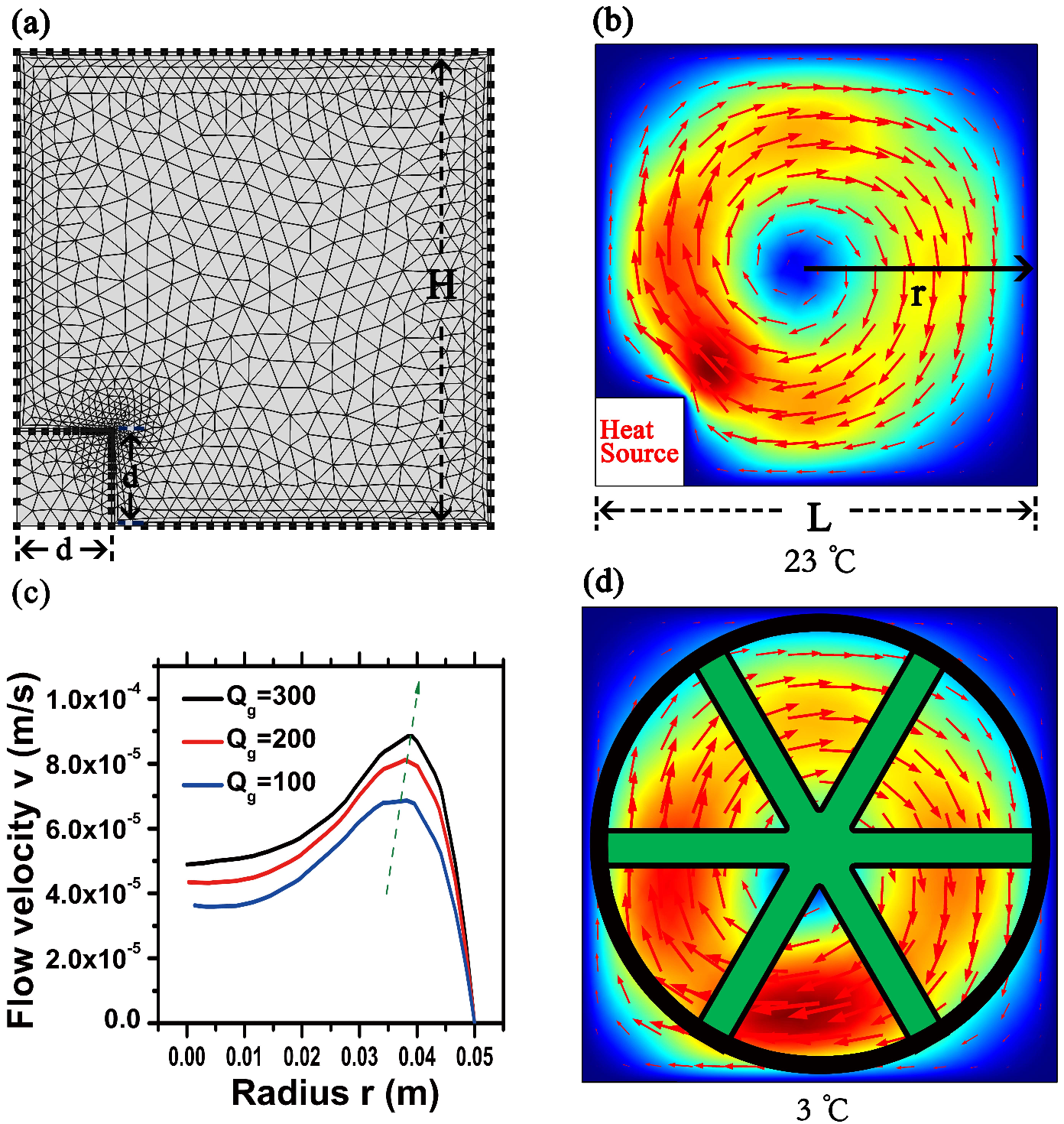}\\
  \caption{The numerically-generated grid system and results of water recirculating flows in $3D$ enclosures. $(a)$ finite-element grid; $(b)$ the vector plot of the recirculating flow; $(c)$ the uprising flow velocity versus the radius measured from the center of the recirculation to the left wall, under various strengths of heat generation. $(d)$ A fictitious hydraulic machine consisting of $6$ slab spokes situated in a cold-bottom-hot-top fluid ambience.}\label{PIC10}
\end{figure}
The feasibility of harvesting oceanic buoyancy-driven hydraulic energy is preliminarily explored. The focus is on two aspects only: $(a)$ the relationship between the thermal energy released from a stationary source at the left bottom corner of an enclosure, $Q_{g}$,  and the rising flow velocity inside the enclosure, $v$; $(b)$ the heat-conduction energy released from the tip of the slab spoke into the water. Both dynamic similarity and geometric similarity between the model and the ocean are beyond the scope of this exploration. Figure $\ref{PIC10}$ shows quasi-steady-state results pertaining to recirculating water flows in a tank with $0.1\,m\times0.1\,m\times0.2\,m$ as width, height, and depth, respectively. In Fig. $\ref{PIC10}(a)$, the finite-element grid is numerically generated. Fine resolution is required near the corner of the heat source ($d=\,0.02\,m$) to avoid the solution divergence. Figure $\ref{PIC10}(b)$ depicts the vector plot of the flow velocity. If the distance, $r$, is measured from the center of the enclosure horizontally toward the right wall, Fig. $\ref{PIC10}(c)$ can be drawn for $|v|$ versus $r$ parameterized in $Q_{g}$. For $Q_{g}=300\,W/m^{3}$, $|v_{max}|$ reaches $8\times10^{-5}\,m/s$ or slightly less than $0.01\,cm/s$. Since the heat source situated at the corner obstructs the flow slightly, the flow velocity at the center of the enclosure is non-zero. Although the magnitude of this momentum appears nearly negligible, the onset of the flow circulation does encourage the possibility that the buoyancy force is capable of overcoming the viscous force, which is the only opposing force to the former. Furthermore, this small velocity can be magnified, in principle, by gear-to-gear transformation, to possibly drive small electricity generators. It is equivalent to a system-flipping frequency of $0.001\,s^{-1}$. From Fig. $\ref{PIC7}(a)$, the conductivity augmentation ratio can be found to be approximately $50$. Hence, with a reasonable set of data for a fictitious machine (Fig. $\ref{PIC10}(d)$), such as $\kappa=15$, cross-sectional area of the slab spoke $=0.01\,m$ wide $\times0.2\,m$ deep, and $L=0.1\,m$, the heat-conduction volumetric rate released from the tip of the spoke is found to be approximately $300\,W/m^{3}$. If the material of the wheel is selected such that its density is close to the density of the water, the wheel may behave like the water itself (but not a foreign object), and a small amount of heating at the bottom may be sufficient to turn the wheel.
\section{Conclusion}
\label{Conclusion}
In energy-system cycles, transient behaviors can be manipulated such that effective thermal properties, including the thermal conductivity and the heat capacity, vary from their intrinsic material-property values to a few orders of magnitude higher. Whenever an energy system is immersed in two or more reservoirs alternatingly, a quantity named ``nonlinear thermal bias'' appears, and can be used to conveniently compute the heat transfer between the reservoir and the system during a cycle and to help set thermostat values. The possibility of harnessing the energy from a fluid system with cold-bottom-hot-top temperature gradients is explored. Both the second law of thermodynamics and Hamiltonian oscillators are additionally considered to prepare links with other related disciplines via studies of transient behaviors.
\section*{Acknowledgments}
\label{Acknowledgments}
This work is supported in part by the 863 project of China under Grant 2013AA03A107, Major Science and Technology Project between University-Industry Cooperation in Fujian Province under Grant Nos.2011H6025 and 2013H6024, NNSF of China under Grant 11104230, Key Project of Fujian Province under Grant 2012H0039 and the Institute for Complex Adaptive Matter, University of California, Davis, under Grant ICAM-UCD1308291.
\section*{References}
\bibliographystyle{elsarticle-num}
\bibliography{reference}

\providecommand{\noopsort}[1]{}\providecommand{\singleletter}[1]{#1}%
\begin{thebibliography}{10}
\expandafter\ifx\csname url\endcsname\relax
  \def\url#1{\texttt{#1}}\fi
\expandafter\ifx\csname urlprefix\endcsname\relax\def\urlprefix{URL }\fi
\expandafter\ifx\csname href\endcsname\relax
  \def\href#1#2{#2} \def\path#1{#1}\fi

\bibitem{li2004thermal}
B.~Li, L.~Wang, G.~Casati, Thermal diode: Rectification of heat flux, Phys.
  Rev. Lett.

\bibitem{kobayashi2009oxide}
W.~Kobayashi, Y.~Teraoka, I.~Terasaki, An oxide thermal rectifier, Applied
  Physics Letters 95~(17) (2009) 171905.

\bibitem{chang2006solid}
C.~Chang, D.~Okawa, A.~Majumdar, A.~Zettl, Solid-state thermal rectifier,
  Science 314~(5802) (2006) 1121--1124.

\bibitem{mahan2008thermoelectric}
G.~Mahan, B.~Sales, J.~Sharp, Thermoelectric materials: New approaches to an
  old problem, Physics Today 50~(3) (2008) 42--47.

\bibitem{gou2010modeling}
X.~Gou, H.~Xiao, S.~Yang, Modeling, experimental study and optimization on
  low-temperature waste heat thermoelectric generator system, Applied energy
  87~(10) (2010) 3131--3136.

\bibitem{majumdar2004thermoelectricity}
A.~Majumdar, Thermoelectricity in semiconductor nanostructures, Science
  303~(5659) (2004) 777--778.

\bibitem{nelson2002organic}
J.~Nelson, Organic photovoltaic films, Current Opinion in Solid State and
  Materials Science 6~(1) (2002) 87--95.

\bibitem{chow2010review}
T.~Chow, A review on photovoltaic/thermal hybrid solar technology, Applied
  Energy 87~(2) (2010) 365--379.

\bibitem{li2005high}
G.~Li, V.~Shrotriya, J.~Huang, Y.~Yao, T.~Moriarty, K.~Emery, Y.~Yang,
  High-efficiency solution processable polymer photovoltaic cells by
  self-organization of polymer blends, Nature materials 4~(11) (2005) 864--868.

\bibitem{schubert2005light}
E.~F. Schubert, T.~Gessmann, J.~K. Kim, Light emitting diodes, Wiley Online
  Library, 2005.

\bibitem{gessmann2004high}
T.~Gessmann, E.~Schubert, High-efficiency algainp light-emitting diodes for
  solid-state lighting applications, Journal of applied physics 95~(5) (2004)
  2203--2216.

\bibitem{che2000thermal}
J.~Che, T.~Cagin, W.~A. Goddard~III, Thermal conductivity of carbon nanotubes,
  Nanotechnology 11~(2) (2000) 65.

\bibitem{spearing2000materials}
S.~Spearing, Materials issues in microelectromechanical systems (mems), Acta
  Materialia 48~(1) (2000) 179--196.

\bibitem{ho1998micro}
C.-M. Ho, Y.-C. Tai, Micro-electro-mechanical-systems (mems) and fluid flows,
  Annual Review of Fluid Mechanics 30~(1) (1998) 579--612.

\bibitem{tao2001numerical}
W.-q. Tao, Numerical heat transfer, Xi¡¯an Jiaotong University Press, Xi¡¯an
  (2001) 430--447.

\bibitem{petrova2006temperature}
H.~Petrova, J.~P. Juste, I.~Pastoriza-Santos, G.~V. Hartland, L.~M.
  Liz-Marz{\'a}n, P.~Mulvaney, On the temperature stability of gold nanorods:
  comparison between thermal and ultrafast laser-induced heating, Physical
  Chemistry Chemical Physics 8~(7) (2006) 814--821.

\bibitem{glezer1997ultrafast}
E.~N. Glezer, E.~Mazur, Ultrafast-laser driven micro-explosions in transparent
  materials, Applied Physics Letters 71~(7) (1997) 882--884.

\bibitem{chen2006semiclassical}
J.~Chen, D.~Tzou, J.~Beraun, A semiclassical two-temperature model for
  ultrafast laser heating, International Journal of Heat and Mass Transfer
  49~(1) (2006) 307--316.

\bibitem{chang2008breakdown}
C.-W. Chang, D.~Okawa, H.~Garcia, A.~Majumdar, A.~Zettl, Breakdown of
  fourier¡¯s law in nanotube thermal conductors, Physical review letters
  101~(7) (2008) 075903.

\bibitem{cao2007equation}
B.-Y. Cao, Z.-Y. Guo, Equation of motion of a phonon gas and non-fourier heat
  conduction, Journal of Applied Physics 102~(5).

\bibitem{shiomi2006non}
J.~Shiomi, S.~Maruyama, Non-fourier heat conduction in a single-walled carbon
  nanotube: Classical molecular dynamics simulations, Physical Review B 73~(20)
  (2006) 205420.

\bibitem{das2003temperature}
S.~K. Das, N.~Putra, P.~Thiesen, W.~Roetzel, Temperature dependence of thermal
  conductivity enhancement for nanofluids, Journal of Heat Transfer 125~(4)
  (2003) 567--574.

\bibitem{mintsa2009new}
H.~A. Mintsa, G.~Roy, C.~T. Nguyen, D.~Doucet, New temperature dependent
  thermal conductivity data for water-based nanofluids, International Journal
  of Thermal Sciences 48~(2) (2009) 363--371.

\bibitem{lepri2003thermal}
S.~Lepri, R.~Livi, A.~Politi, Thermal conduction in classical low-dimensional
  lattices, Physics Reports 377~(1) (2003) 1--80.

\bibitem{dhar2008heat}
A.~Dhar, Heat transport in low-dimensional systems, Advances in Physics 57~(5)
  (2008) 457--537.

\bibitem{hatano2001steady}
T.~Hatano, S.-i. Sasa, Steady-state thermodynamics of langevin systems,
  Physical review letters 86~(16) (2001) 3463.

\bibitem{sekimoto1997kinetic}
K.~Sekimoto, Kinetic characterization of heat bath and the energetics of
  thermal ratchet models, Journal of the physical society of Japan 66~(5)
  (1997) 1234--1237.

\bibitem{21honeycutt1992stochastic}
R.~L. Honeycutt, Stochastic runge-kutta algorithms. i. white noise, Physical
  Review A 45 (1992) 600--603.

\bibitem{hull1972comparing}
T.~Hull, W.~Enright, B.~Fellen, A.~Sedgwick, Comparing numerical methods for
  ordinary differential equations, SIAM Journal on Numerical Analysis 9~(4)
  (1972) 603--637.

\bibitem{comsol2005comsol}
A.~Comsol, Comsol multiphysics user¡¯s guide, Version: September.

\bibitem{shih1984numerical}
T.~M. Shih, Numerical heat transfer, Springer-Verlag, New York, 1984.

\end{thebibliography}
\appendix
\begin{appendix}
\section{}\label{}
\subsection{Derivation of Eq.(1)}\label{Derivation of Eq.(1)}
When $\kappa$ depends on $T$, the partial differential equation governing $1D$ transient heat conduction can be written as
\begin{equation}\label{A.1}
\rho c_{v}\frac{\partial T}{\partial t}=\frac{\partial}{\partial x}(\kappa\frac{\partial T}{\partial x})+Q_{g},
\end{equation}
which can be rearranged into
\begin{equation}
\rho c_{v}\frac{\partial T}{\partial t}=\kappa\frac{\partial^{2}T}{\partial x^{2}}+\frac{\partial\kappa}{\partial x}\frac{\partial T}{\partial x}+Q_{g}\\\nonumber
\end{equation}
or
\begin{equation}\label{A.2}
\rho c_{v}\frac{\partial T}{\partial t}=\kappa\frac{\partial^{2}T}{\partial x^{2}}+(\frac{\partial\kappa}{\partial T})(\frac{\partial T}{\partial x})^{2}+Q_{g}.
\end{equation}
In bi-segment thermal rectifiers, the left segment in contact with the hot reservoir exhibits the character of $\partial\kappa/\partial T>0$, thus rendering the second term in the right-hand side of Eq. (\ref{A.2}) positive. A positive term behaves like a heat source, inducing a large temperature gradient near the bi-segment junction. In the right segment, however, the opposite character, namely $\partial\kappa/\partial T<0$, yields a small temperature gradient near the junction. Therefore, a thermal rectification effect prevails. The derivation for the $3D$ equation is similar to that for $1D$.
\subsection{Derivation of Eq.(12)}\label{Derivation of Eq.(12)}
If $T_{\alpha}$ is set to $T_{i}$, $t_{qs}$ must be equal to zero in Eq. (\ref{10}). Adding Eqs. (\ref{10}) and (\ref{11}) yields
\begin{equation}\label{A.3}
T_{\alpha}+T_{\beta}=T_{Rh}+T_{Rc}=2T_{Rh}-(T_{Rh}-T_{\alpha})[1+exp(-c_{1}t_{o})],
\end{equation}
which can be readily simplified to Eq.(\ref{12}). As a result, $T_{\alpha}$ can be expressed analytically in terms of all prescribed parameters, and $T_{\beta}$ can be determined simply by
\begin{equation}\label{A.4}
T_{\beta}=T_{Rh}+T_{Rc}-T_{\alpha}.
\end{equation}
\subsection{Derivation of Eq.(16)}\label{Derivation of Eq.(16)}
By definition, $r_{cond}=J_{qs}/J_{ss}$. When $f$ approached infinity, $T_{\alpha}$ and $T_{\beta}$ merge into $T_{m}=(T_{Rh}+T_{Rc})/2$. Therefore, in conjunction with Eq.(\ref{15}), this definition leads to\\
\begin{equation}
\begin{split}
r_{cond}&=\frac{hA(T_{Rh}-T_{m})}{hA(T_{Rh}-T_{ss})}=\frac{0.5(T_{Rh}-T_{Rc})}{T_{Rh}-T_{ss}}\\
&=\frac{0.5(T_{Rh}-T_{Rc})(B+2)}{(B+2)T_{Rh}-(B+1)T_{Rh}-T_{Rc}},
\end{split}
\end{equation}
which yields Eq.(16) after straightforward algebra.
\subsection{Derivation of $3D$ transient heat conduction equations}\label{Derivation of $3D$ transient heat conduction equations}
Resulting algebraic equations governing $T(1)$, $T(2)$, $\cdots$, $T(6)$ can be derived based on the energy conservation over individual cubes as
\begin{equation}
\footnotesize
\left(
  \begin{array}{cccccc}
    1  &      &            &      &      &  \\
    -r & 3r+1 & -r         & -r   &      &  \\
       & -r   & r+rb_{i}+1 &      &      &  \\
       & -r   &            & 2r+1 & -r   &  \\
       &      &            & -r   & 3r+1 & -r \\
       &      &            &      & -1   & 1 \\
  \end{array}
\right)[T]=\left(
           \begin{array}{c}
             T(1) \\
             T^{p}(2) \\
             T^{p}(3)+rb_{i}T_{\infty} \\
             T^{p}(4) \\
             T^{p}(5)+rT(7) \\
             0 \\
           \end{array}
         \right).
\end{equation}
\nomenclature{$b_{i}$}{small Biot number defined as $h\Delta x/\kappa$ (dimensionless)}
\nomenclature{$r$}{aspect ratio, $\alpha \Delta t/(\Delta x)^{2}$ (dimensionless)}
Parameters are defined as $r=\alpha\Delta t/(\Delta x)^{2}$ and $b_{i}=h\Delta x/k$; $\Delta x=0.01\,m$; and $\Delta t=10\,s$. The temperature of cube $1$ is intentionally treated as an unknown for indexing convenience.
\nomenclature[G]{$\alpha$}{thermal diffusivity, $\kappa/(\rho c_{v})$ ($m^{2}/s$)}
\subsection{Analytical solution $T(t)$ when $nx=1$ subject to continuously-changing boundary conditions}\label{Analytical solution $T(t)$ when $nx=1$ subject to continuously-changing boundary conditions}
Temperatures of hot and cold reservoirs are prescribed as $A_{o}\cos(ft)$ and $-A_{o}\cos(ft)$, respectively. The initial condition is $T_{1}(0)=T_{Rh}$, $T_{2}(0)=T_{Rc}$. Governing equations for this system can be written as
\begin{equation}
\frac{d T_{1}(t)}{dt}=\frac{h}{\rho c_{v}}\left[A_{o}\cos{(ft)}-T_{1}(t)\right]-\frac{\kappa}{\rho c_{v}L}\left[T_{1}(t)-T_{2}(t)\right] \nonumber
\end{equation}
and
\begin{equation}
\frac{d T_{2}(t)}{dt}=-\frac{h}{\rho c_{v}}\left[A_{o}\cos{(ft)}+T_{2}(t)\right]+\frac{\kappa}{\rho c_{v}L}\left[T_{1}(t)-T_{2}(t)\right]\nonumber,
\end{equation}
which can be solved analytically to yield
\begin{equation}
T_{1}(t)=\frac{g_{5}+g_{1}(1-g_{2})+g_{2}g_{4}[(g_{3}+1)T_{Rh}+(g_{3}-1)T_{Rc}]}{2g_{4}},
\end{equation}
where\\
\begin{equation}
g_{1}=2A_{o}hL(2\kappa+hL)\cos(ft)\nonumber,
\end{equation}
\begin{equation}
g_{2}=\exp(-(2\kappa t+hLt)/(\rho c_{v}L))\nonumber,
\end{equation}
\begin{equation}
g_{3}=\exp((2\kappa t)/(\rho c_{v}L))\nonumber,
\end{equation}
\begin{equation}
g_{4}=4\kappa^{2}+4hL\kappa+L^{2}(h^{2}+\rho^{2}c_{v}^{2}f^{2})\nonumber,
\end{equation}
and
\begin{equation}
g_{5}=2A_{o}c_{v}\rho hL^{2}f\sin(ft)\nonumber.
\end{equation}
The solution for $T_{2}(t)$ is omitted. Figure \ref{PIC9} shows $r_{cond}$ versus $f$ parameterized in $h$.
\end{appendix}


\end{document}